\documentclass[sigconf]{acmart}
\usepackage{titlesec}
\usepackage{float}

\titlespacing{\section}{0pt}{4pt}{0pt}
\titlespacing{\subsection}{0pt}{4pt}{0pt}
\AtBeginDocument{%
  \providecommand\BibTeX{{%
    \normalfont B\kern-0.5em{\scshape i\kern-0.25em b}\kern-0.8em\TeX}}}

\copyrightyear{}
\acmYear{}
\acmDOI{}

\setcopyright{none}

\acmConference[KDD IRS2020]{International Workshop on Industrial Recommendation Systems 2020}{Aug 23--28, 2020}{San Diego, CA}
\acmISBN{}

\usepackage{lipsum}
\usepackage{svg}
\usepackage{makecell}
\usepackage{paralist}

\begin{document}

\title{Lessons Learned Addressing Dataset Bias in Model-Based Candidate Generation at Twitter}

\author{Alim Virani*, Jay Baxter*, Dan Shiebler*, Philip Gautier, Shivam Verma, Yan Xia, Apoorv Sharma, Sumit Binnani, Linlin Chen, Chenguang Yu}
\email{[avirani,jbaxter,dshiebler,pgautier,shivamv,yanx,apoorvs,sbinnani,linlinc,chenguangy]@twitter.com}
\affiliation{%
  \institution{Twitter Cortex}
  \thanks{* Primary Contributors}
}

\begin{abstract}
  Traditionally, heuristic methods are used to generate candidates for large scale recommender systems. Model-based candidate generation promises multiple potential advantages, primarily that we can explicitly optimize the same objective as the downstream ranking model. However, large scale model-based candidate generation approaches suffer from dataset bias problems caused by the infeasibility of obtaining representative data on very irrelevant candidates. Popular techniques to correct dataset bias, such as inverse propensity scoring, do not work well in the context of candidate generation. We first explore the dynamics of the dataset bias problem and then demonstrate how to use random sampling techniques to mitigate it. Finally, in a novel application of fine-tuning, we show performance gains when applying our candidate generation system to Twitter's home timeline. 
\end{abstract}




\maketitle

\section{Introduction}
\begin{figure}[hbt]
	\includegraphics[width=\linewidth]{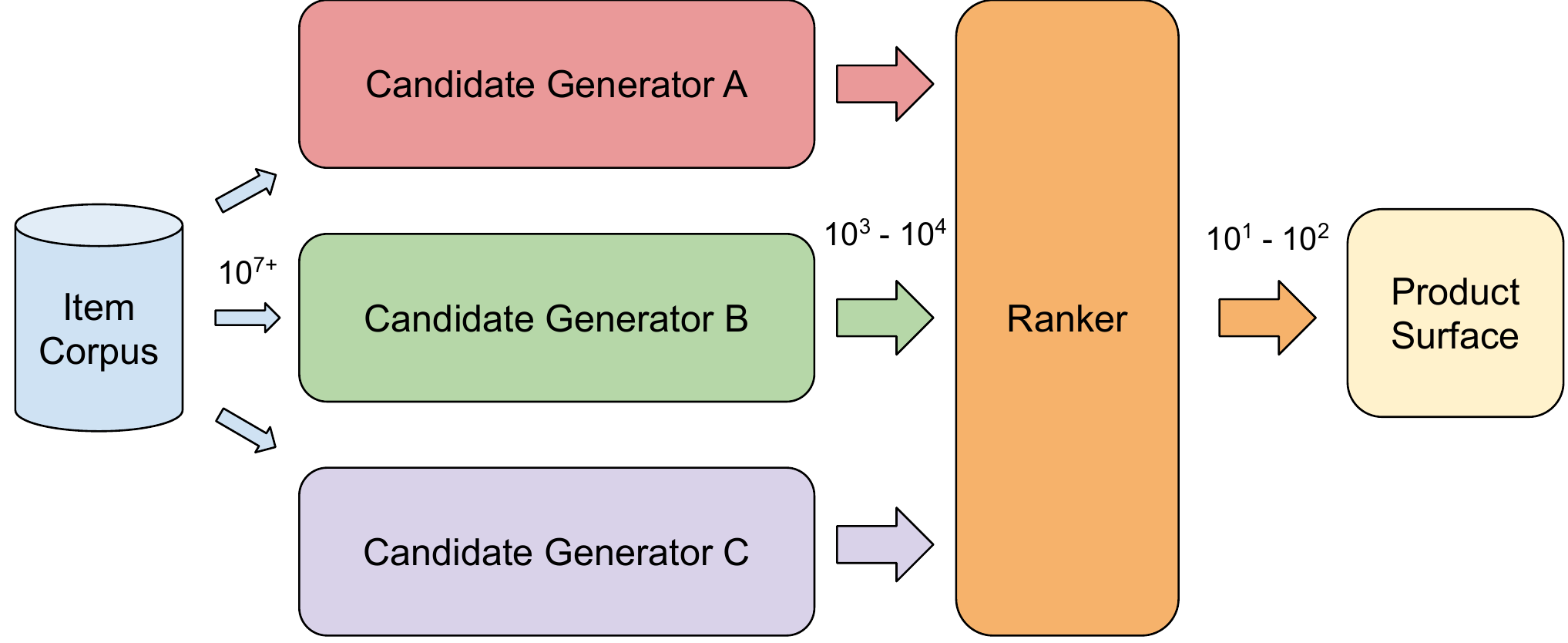}
	\caption{Candidate generation and ranking pipeline.}
	\label{fig:pipeline}
\end{figure}

Most recommender systems at Twitter operate at a massive scale. These systems recommend to our users Tweets they will engage with, ads they might click on, and other users they might follow. Often the corpus of candidates is too large to exhaustively score all items for every user using the heavy ranker. Rather, when operating at this scale, recommender systems go through a candidate generation (CG) phase (see Figure~\ref{fig:pipeline}) which  efficiently narrows the entire corpus down to a manageable candidate set. After this phase, the retrieved candidates are then ranked using a computationally heavier model. 

Because scoring all candidates is infeasible, large-scale CG systems at Twitter have relied on a combination of domain knowledge, heuristics and handwritten rules to narrow down the candidate set. For example, if a user has engaged with the content of a given author in the past, a CG algorithm could suggest new content from the same author.

While heuristic CG systems have been highly successful in the past, they can be difficult to extend and maintain, which leads to performance plateaus. 
Additionally, heuristic candidate generation methods usually lack the model capacity to represent the more complex relationships, resulting in lower quality recommendations.

For instance, the individual heuristic CG algorithms that source Tweets for a user's timeline may recommend the top Tweets in that user's geographical area and the top Tweets in that user's interest areas separately. The union of these candidate sources may exclude a Tweet that was barely excluded by both heuristics but which is relevant and engaging to the user because of its multi-faceted appeal.



Model-based candidate generation is one potential solution to these problems, but it requires careful handling of several practical challenges before outperforming heuristics in production. This paper discusses some of the practical challenges that arise in building such a system.

The first challenge is scale: such a model would need to generate candidates for a user from the entire corpus of items with very low latency.  To achieve this performance, we use a two tower network to learn query-item co-embeddings that we then serve with an approximate nearest neighbors (ANN) system.

Another major challenge, and the focus of this paper, is dataset bias. The theoretical challenges of dataset biases are well-documented in the literature \cite{DBLP:journals/corr/abs-1812-02353} \cite{DBLP:journals/corr/JoachimsSS16}. However, most of the existing literature focuses on the dataset bias problem as it relates to the ranking phase of recommender systems, whereas our focus is specifically on candidate generation. Datasets created by sampling from served data are not representative of the entire corpus, and as a result, models trained on these datasets have been shown to create filter bubbles \cite{filterbubble} and degenerate feedback loops \cite{Jiang_2019}. The impact of bias in CG is particularly challenging because in CG, the inference set is the entire corpus and the cost of serving very low-quality candidates is prohibitively high.


This work is an account of the lessons we have learned when dealing with bias in data from served traffic in the context of model based candidate generation. The contributions of this paper include:
\begin{compactitem}
  \item Empirical evidence for the impact of dataset bias on a model-based candidate generation system.
  \item A set of practical suggestions for how to do random negative sampling in such a system.
  \item A novel application of fine-tuning to training a two tower model for candidate generation. This approach shows significant wins in a large scale online A/B test making Twitter home timeline recommendations.
\end{compactitem}

The rest of the paper is structured as follows: We begin in Section~\ref{problem} by giving a description of the dataset bias problem in model-based CG and some shortcomings in traditional dataset bias correction approaches such as inverse propensity score weighting and naive random negative sampling. In Section~\ref{setup}, we describe our model, system architecture, and our model evaluation criteria. In Section~\ref{solutions}, we describe techniques for improving random negative sampling and show results demonstrating their efficacy. Finally, in Section~\ref{fine-tuning}, we describe the fine-tuning approach to bias correction and include its successful results in production.

\section{Related Work}

\subsection{Dataset Bias}

There are multiple sources of data-related biases that occur across the recommendation pipeline. One example is position bias \cite{Hofmann2014EffectsOP} which is caused by the order in which items are displayed. Another is popularity bias \cite{popularity19} which is caused by recommending some items in the corpus far more frequently, causing negative effects such as the inability to suggest less popular (long tail) items \cite{sloan06}. 

The dataset bias central to this work, a form of selection bias, is caused by only observing feedback on recommendations made from existing recommender systems. 

One approach attempts to correct for this bias by intervening in data collection through randomized exploration \cite{DBLP:journals/corr/JoachimsSS16}. In particular, Chen et al. \cite{DBLP:journals/corr/abs-1812-02353} address dataset bias in candidate generation by promoting off-policy exploration using techniques from reinforcement learning. Our approach does not attempt to correct for bias using exploration due to product constraints which prevent us from using the same exploration techniques.

Swaminathan et al. \cite{DBLP:journals/corr/SwaminathanJ15} uses an approach to correct for bias that does not require active exploration. Rather, they use counterfactual estimates of performance and inverse propensity scoring (IPS). However, their approach is limited to the ranking phase. Indeed, they point out that this bias is particularly difficult in the context of candidate generation: Not only are some recommendations over-represented but others are missing altogether.

\subsection{Two-Tower Networks} 


Two-tower networks, as described in Section~\ref{twotower}, are a class of deep learning models useful for co-embedding queries and candidates in a common embedding space. Query and candidate embeddings are computed separately by each tower. This avoids having to recompute candidate embeddings for every query. Importantly, it allows those embeddings to be indexed into an efficient approximate nearest neighbor system. 

This makes them a popular choice as a candidate generation model \cite{krichene2018efficient, Yi:2019:SNM:3298689.3346996}. Covington et al. \cite{Covington2016DeepNN} use a two-tower architecture to co-embed YouTube users and YouTube videos.

Negative sampling is a common technique in recommender systems, especially in the context of sampling from implicit negative feedback. In such cases, model performance is highly sensitive to the sampling distribution and the number of sampled negatives \cite{DBLP:journals/corr/abs-1708-05031}. Our findings show that this extends to the context of model-based candidate generation. 

\subsection{Approximate Nearest Neighbor Search} \label{ANN}

Exact nearest neighbor algorithms do not work well on relatively high dimensional data: due to the curse of dimensionality, it is infeasible to find an exact K-nearest neighbor solution when dealing with a massive corpus. A more practical solution is to use an approximate nearest neighbor search which allow some errors in return for much higher search speed. 

There are many open source implementations of approximate nearest neighbor search. Examples include Spotify's Annoy \cite{annoy} and Facebook's Faiss \cite{faiss}. Our framework used the graph-based HNSW algorithm \cite{DBLP:journals/corr/MalkovY16}. HNSW offers state-of-the-art performance \cite{DBLP:journals/corr/abs-1807-05614} and supports index updates.

\section{Problem Description} \label{problem}

In the following section we study the dataset bias problem and the use of sampled random negatives. This will motivate our contributions in Section~\ref{solutions} and \ref{fine-tuning}. 


Let us define a recommendation query $q$ as a request for a recommendation for a user in a particular context, where a context captures all known information about the user's request (e.g. the time of day, the current state of the user's app, etc.). A ``candidate'' is an element of the set of items that we want to recommend. For example, a context could be a user opening the Twitter app by clicking on a particular notification (and would include information about that notification), and our set of candidates could be the set of all recent Tweets.

Consider the corpus of all candidates  $c_1, c_2 \dots c_n \in C$ and the set of all queries $q_1, q_2 \dots q_m \in Q$. For a given query $q_j$ and candidate $c_i$ served to the user, the binary label $r_{ij}=1$ if the user engages with (e.g. favorites) the candidate. 
Because the relevance depends on user feedback, we model it as stochastic, i.e. $r_{ij}=1$ with probability $p_{ij}$ and 0 otherwise. 


The role of candidate generation is to return a candidate set $C_q$ for a query $q$ that is a subset of the corpus $C$ such that $C_q$ contains the candidates with the highest $p_{ij}$, i.e. the best candidates to serve for this query. 
We infer a model $f$ which takes in a (query, candidate) pair and returns the probability that the candidate is relevant to the query, i.e. $f(c_i,q_j) = p_{ij}$. 


For a given query $q_j$ we can partition the corpus into three broad categories. The first two categories contain Tweets that are irrelevant to the query in the sense that $p_{ij} \approx 0$. They are differentiated, however, by feedback external to the relevance problem: whether they are a detriment to the user experience beyond the scope of the query. 

\textbf{Relevant and engaging candidates}: First, there are relevant candidates, where for many of these candidates $p_{ij}$ is high.


\textbf{Relevant but not engaging candidates}: These candidates are relevant to the query and don't provoke a strong negative reaction. However, they also aren't engaging: $p_{ij}$ is close 0. As such their contribution to user experience is limited, but at least the impact to the user experience is not particularly negative beyond a potentially boring interaction.

\textbf{Extremely irrelevant candidates}: Finally, extremely irrelevant recommendations, such as showing a user a Tweet in a language they do not understand, may frustrate the user and deteriorate their faith in the platform. Some candidates are not only irrelevant, but also are harmful to the user experience and directly harm retention. We note that language mismatch is merely an illustrative example and does not cover all possible reasons a recommendation could be in this category.

 We can decompose the candidate generation task into the following two components, which we address separately:  
\begin{compactitem}
  \item Avoid retrieving "extremely irrelevant" candidates
  \item Retrieve "relevant and engaging" candidates rather than "relevant but not engaging" candidates
\end{compactitem}

In the remainder of this section, we will show that naively training on served traffic attempts to solve the second goal but not the first, and training only on randomly sampled negatives solves the first goal but not the second.

\subsection{Served Traffic} \label{served}

A standard formulation for constructing a dataset $D$ with which to to train $f$ is to randomly sample $(c_i, q_j, r_{ij}) \in D$ i.i.d from a uniform random distribution over the set of all candidates $C$ and all queries $Q$. In order to construct such a dataset, we would need to serve truly random candidates to users. Unfortunately, this is not a realistic option because we will serve "extremely irrelevant" candidates to users. If a user in Germany opened Twitter tomorrow morning and saw a mishmash of random Japanese and Spanish Tweets on their Timeline, this would almost certainly be an unacceptably poor experience for that user.

In lieu of serving with a uniform probability  we could instead serve according to a probability weighted by $p_{ij}$. If we serve (query, candidate) pairs according to a distribution that approximated $p_{ij}$ then we avoid serving mostly "extremely irrelevant" candidates to users. In principle, as long as $Q(q_j, c_i) > 0$ for all $i$ and $j$, then we could use inverse propensity score weighting to weight training examples. While we can not observe $p_{ij}$  we can use an existing model that approximate it. 


In practice, importance sampling is challenging in the context of candidate generation for two reasons. First, existing models do not perform well across entire corpus and therefore cannot be used to guide sampling. More importantly, it would still require us to show a nontrivial amount of "extremely irrelevant" candidates that will actively hurt user experience, which is not acceptable from a product perspective.  

In practice we can only draw labeled training examples from served candidates $C_q^s$ that have been selected by some existing candidate generators.

\begin{figure}[hbt]
    \centering
    \includegraphics[width=\columnwidth]{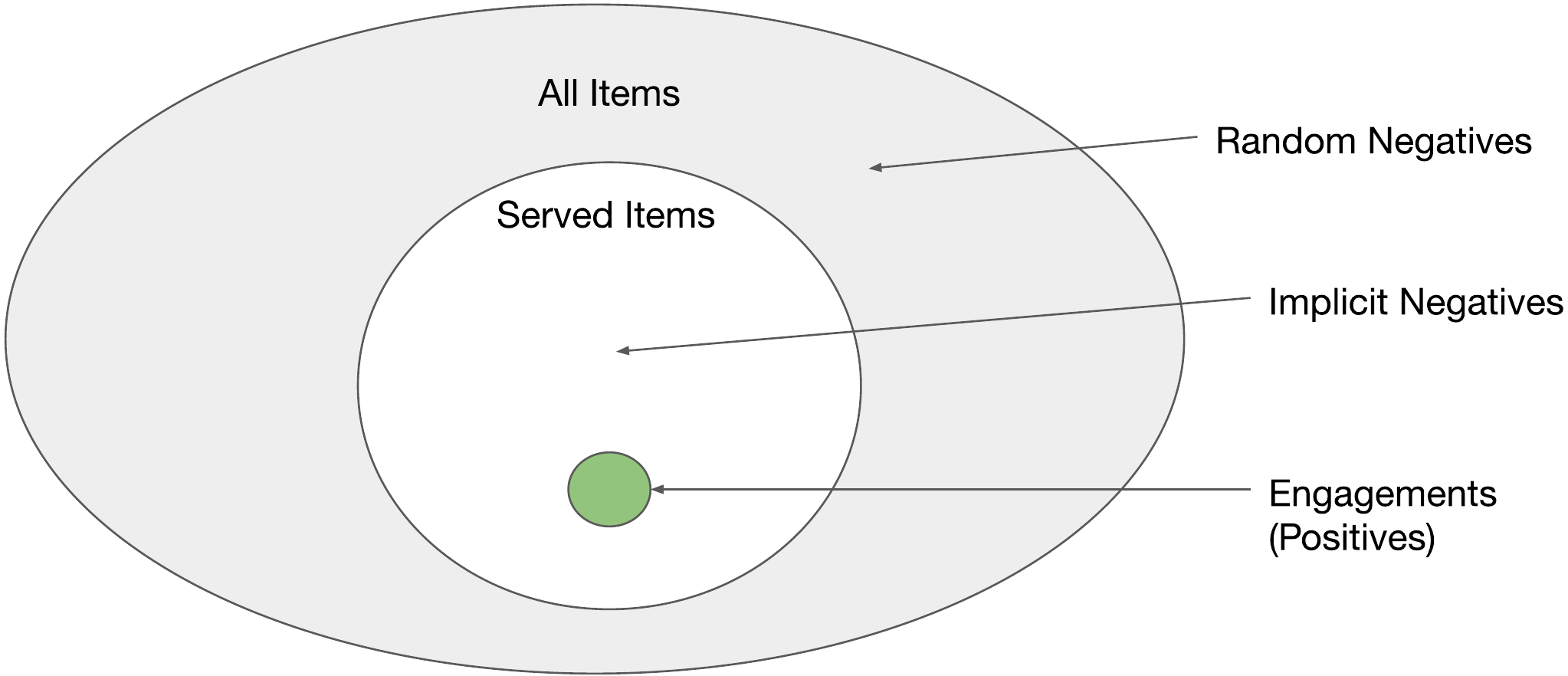}
    \caption{These concentric circles (not to scale) illustrate the three categories of datasets that we discuss in Sections~\ref{expimp} and \ref{samplednegs}.
    \label{fig:concentric_circles}}
\end{figure}

\subsection{Explicit and Implicit Feedback}\label{expimp}

For most product surfaces at Twitter, we can get consistent and high quality positive feedback for relevant query candidate pairs. When a user engages with a candidate (e.g. clicking on an item) she is explicitly indicating that this candidate is relevant to her. When training the two-tower model we derive positive (query, candidate) pairs (i.e. $r_{ij}=1$) from these actions.


Unlike positive examples, it is very challenging to derive negative examples from explicit negative feedback.  Unfortunately, very few product surfaces across Twitter get significant amounts of explicit negative feedback from users. Such feedback tends to be minimal and highly skewed towards very specific circumstances.  For instance, only a small percentage of our users ever use the "show less often", mute or block buttons available on their Tweet timelines, since they require more than one click.

As an alternative we can obtain irrelevant (query, candidate) pairs using implicit negative feedback. If a user does not follow an account we suggested, she implicitly tells us that this suggested account was not relevant to her.

Clearly the lack of positive engagement cannot always be understood to mean a lack of relevance. There are many reasons why a user may not engage with a recommended candidate that are not related to its relevance, so implicit negative feedback will always be noisy. Nevertheless, we can assume that these implicit negatives are less relevant on average than those items the user has given positive feedback for. 

Figure \ref{fig:concentric_circles} illustrates these categories: engagements (positives) are explicit feedback and are all "relevant and engaging", served but not engaged items are implicit negative feedback (mostly "relevant but not engaging"), and random negatives comprise all not served content, which has all three types of content, but mostly "totally irrelevant and not engaging" and "relevant but not engaging".

An intuitive approach is then to construct the dataset by sampling (query, candidate) pairs $(c_i, q_j)$ from $C_{q_{j}}^s$. We set $r_{ij}=1$ if the pair has positive explicit feedback (i.e. it has been engaged with) and low if it has negative implicit feedback (i.e. it has not been engaged with).

Unfortunately, models trained in this way perform very poorly when used for model based candidate generation. The reason for this is that we require our model to find the best candidates in the entire corpus $C$ even if all our training data is derived from $C_q^s$. If the model has never seen any candidates from outside of $C_q^s$, its performance may be nearly random on the vast majority of $C$. The reason for this is that $C_q^s$ contains only (query, candidate) pairs where the candidate was recommended to the query by an existing recommendation algorithm. There are extremely few elements of $C_q^s$ for which the candidate is a terrible recommendation for the query. Shown in Table \ref{tab:offline}, a model trained this way recommends off-language content to a high proportion of users (over 75\%).


\subsection{Sampled Negatives}\label{samplednegs}

Given that training on served traffic alone produces models that make poor recommendations, what else can we resort to?

One thing we can do is take advantage of the property that for a random query $q_j$, $p_{ij} \approx 0$ for most $c_i \in C$. 
Therefore, we can reasonably
treat randomly sampled candidates as negatives.

When we train our model with explicit feedback samples as positives and random samples as negatives, our model learns to distinguish relevant content from completely irrelevant content.
However, a model trained this way will be less effective at distinguishing somewhat relevant content (where $p_{ij}$ still tends to be near 0) from engaging content.

After retrieval, the candidate set $C_q$ will still go through a ranking phase. In that sense, it’s the ranking model's task to be able to distinguish between the "relevant but not engaging" and "relevant and engaging" content in $C_q$. Still, the larger the portion of candidates $c_i$ in $C_q$ that have $p_{ij}\approx 0$, the more the quality of the output ranking will suffer: the number of "relevant and engaging" candidates that can be outputted by the ranker is bounded by the number of "relevant and engaging" candidates included in the candidate generation phase. 

Shown in Table \ref{tab:offline}, these models perform better than those trained on negatives from implicit feedback, since they can correctly filter out "extremely irrelevant" candidates. The challenge of training a candidate generation model using sampled random negatives is to improve performance on the second objective -  retrieving candidates that will be engaging - without sacrificing the former. 

In sections \ref{solutions} and \ref{fine-tuning} we present our solutions to address this challenge.

\subsection{Demonstration of Bias on MovieLens}\label{movielens}

To illustrate the issue of bias in model-based candidate generation, we conduct an experiment on the MovieLens 1M dataset \cite{harper2016movielens}. The experiment simulates the process of only training on biased data that has been selected by a previous ranker. It shows that, even as performance on the biased test set increases, performance on the full test decreases.

First, we interpret all samples in the Movielens dataset as explicit positives. Next, we uniformly sample random (user, movie) pairs and treat these random samples as negative samples. Then, we distinguish these ``explicit positives'' from the random negatives by training an MLP classifier which takes user and movie embeddings as input.

\begin{figure}[H]
	\includegraphics[width=\linewidth]{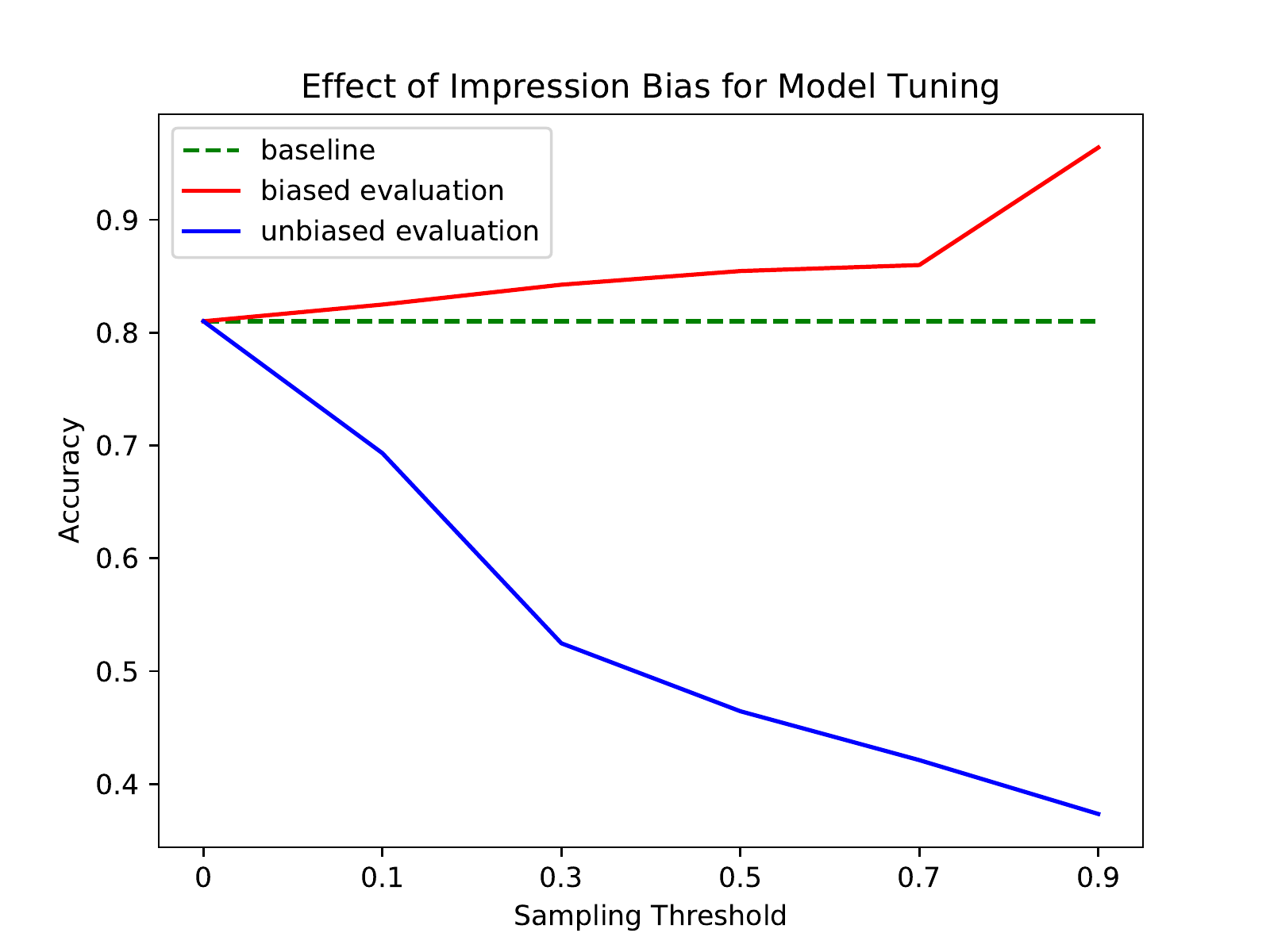}
	\caption{Effect of the biased data for model training: as the sampling threshold $\tau$ increases, performance on the biased test set increases while performance on the full test set decreases.}
	\label{fig:bias_sampling_effect}
\end{figure}

We train this model in two phases. In the first phase we simply train the model on the full dataset. We split the dataset into training/testing based on user ids: we place 80\% of users into the train dataset and 20\% into the test dataset. We report the performance of the model in this stage with the green dotted line in Figure \ref{fig:bias_sampling_effect}.


In the second phase we simulate the effect of training on biased data. First, we run the unbiased model from the first phase to score every sample in the train and test datasets. Then, we vary the sampling ratio $\tau$, where $\tau$-biased train/test datasets are the subsets of the train/test datasets that include only examples with a predicted score from the unbiased model larger than $\tau$.

These biased datasets are akin to datasets derived from only served data: they only include recommendations that a previous model ranked highly.
The red line in Figure \ref{fig:bias_sampling_effect} shows the performance of a model trained on the $\tau$-biased train dataset over the $\tau$-biased test dataset, and the blue line shows the performance of this model on the full test dataset.

As $\tau$ increases, we see that performance on the $\tau$-biased test dataset increases, but performance on the full unbiased dataset decreases.
That is, if we only look at the model's performance on served traffic, the model appears to perform better as we set stronger filters on the data that is ``served in production'', 
but performance on the entire dataset (including the unserved candidates below the $\tau$ threshold) actually decreases. Unfortunately, in the case of candidate generation at Twitter scale, $\tau$ is extremely high as the number of available items to recommend is many orders of magnitude larger than the number served. This provides evidence as to why we observe that a model based candidate generation algorithm trained only on served traffic performs so poorly.

\section{Experimental Setup} \label{setup}

\subsection{Product Surfaces} \label{product}

Because the two tower network model-based candidate generation approach is quite general, in this paper we describe applications to four different Twitter products.

In the "Home Timeline", "Who To Follow", and "News Article Push Recommendations" products the recommended items are Tweets, accounts to follow, and news article links respectively. The "Notification Landing Page" is similar to the home timeline in that we recommend Tweets to users, but has a different context: where the home timeline context is more general, the notification landing page's context includes the specific Tweet the user just clicked in their notification, so we are able to show similar content to that target Tweet.

While further details of the product concerns specific to these applications is outside the scope of this paper, we include results from multiple products to demonstrate the flexibility and generality of this candidate generation approach.

\subsection{Two Tower Networks} \label{twotower}

\begin{figure}[hbt]
	\includegraphics[width=\linewidth]{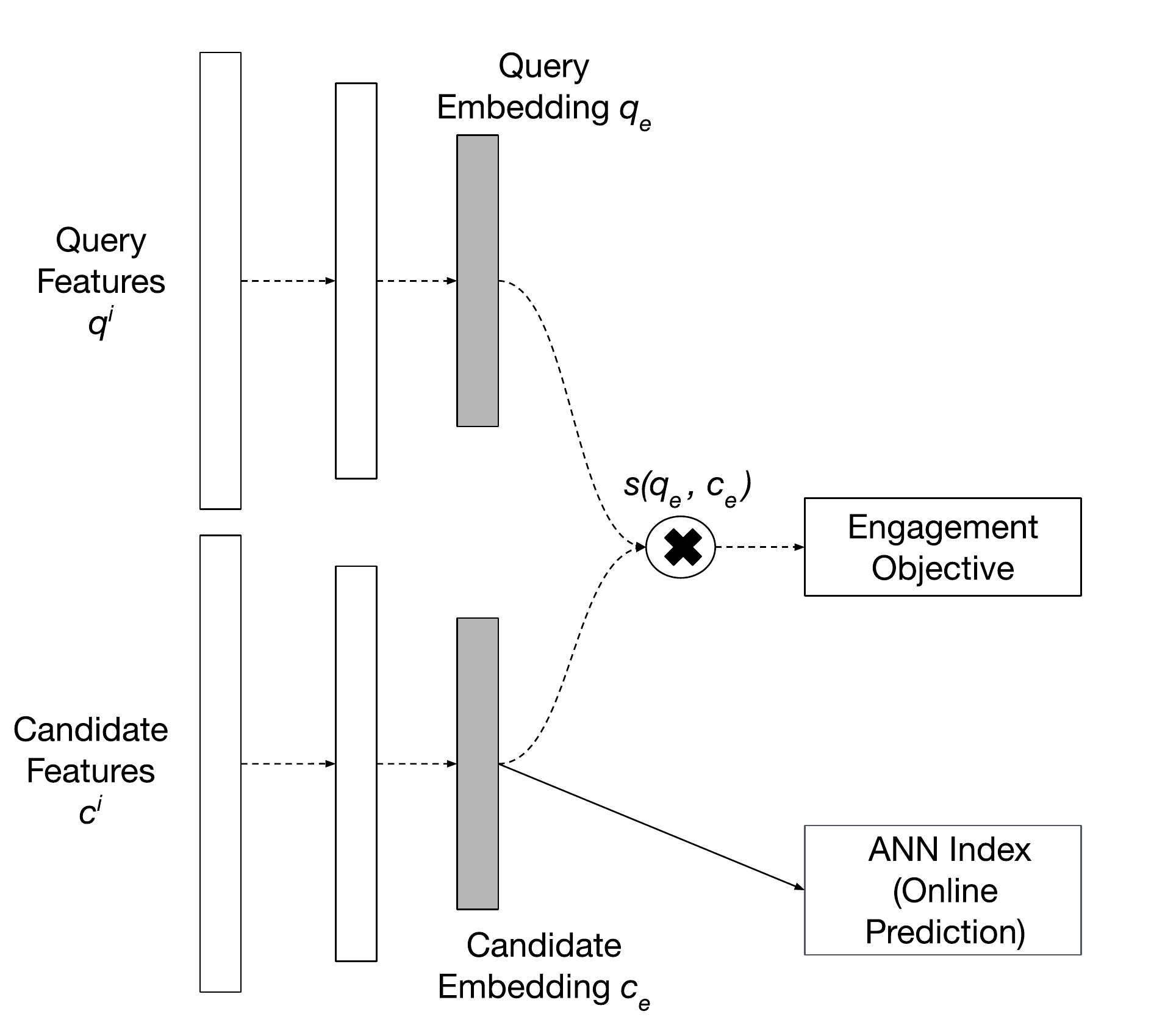}
	\caption{Two Tower Network and ANN.}
	\label{fig:architecture}
\end{figure}

In our approach, for all product surfaces, we use a two tower network (see Figure~\ref{fig:architecture}) to encode query and candidate embedding vectors, $q_e$ and $c_e$ respectively, from input features, $q^{i}$ and $c^{i}$ respectively.

Each tower of the model learns a separate function, $h(c^{i})$ and $g(q^{i})$, which encode the embeddings' vectors: $c_e = h(c^{i})$ and $q_e=g(q^{i})$. These functions are learned together as the towers are trained jointly to co-embed relevant (query, candidate) pairs in close proximity and irrelevant ones further apart. 
The modeling task can be framed as classification where the positive cases are relevant (query, candidate) pairs and negative if they are irrelevant. Such models can be trained using a binary classification loss function like cross entropy.

In this way, we model relevance as proportional to the similarity between a (query, candidate) embedding vector:
$\hat{f} = \hat{p_{ij}} = \sigma(s(c_e, q_e))$  for some similarity function $s$. Cosine similarity is a commonly used metric to calculate the similarity (relevance) between both embeddings (and the one we use in practice throughout the paper's results), but other viable alternatives include the inner product and euclidean distance.

Given a query $q$ and its embedding vector $q_e$, we create a candidate set $C_q$ from $C$ by finding the K candidates with the closest embedding vectors to $q_e$. In a standard K nearest neighbour search, where this model would be evaluated for each candidate $c \in C$ for every new query $q$, these searches would be prohibitively computationally expensive. The reason for the two-tower architecture (rather than fully connected) is that it enables us to decouple the user and candidate networks. While both are trained jointly, when used in production at inference-time, the candidate network is evaluated either at write-time (as candidate features are updated, e.g. a Tweet receives new engagements) or on a fixed cadence, and the resulting candidate embeddings are loaded into the nearest neighbor index. Then, when a time-sensitive query (user-context pair) is made, the nearest neighbor index looks up the closest neighbors $c_e$ to $q_e$ and returns them. In order to compute nearest neighbors efficiently, we use approximate algorithms- namely HNSW \cite{DBLP:journals/corr/MalkovY16}.

This modeling setup offers an intuitive explanation of why we cannot train on served traffic ($C_q^s$): A well formed embedding space would be one that maintained a distance between every (query, candidate) pair in proportion to their relevance, including pairs that cannot be derived from $q$ and $C_q^s$.

\subsection{Model Evaluation}\label{evaluation}

In Covington et Al. \cite{Covington2016DeepNN}, they stress that online A/B testing had an outsized importance when testing their candidate generation models. While we also use A/B testing to get definitive results for a model, we would also like to evaluate the quality of a model offline. Testing models online using A/B tests is a time consuming process and one that is not suitable to guide model iteration. Ideally, we want an offline test that correlates to online performance on engagement metrics and hopefully also user retention metrics. 

Unfortunately, model performance on a single test set is not sufficient to ensure that the candidates retrieved will be relevant and engaging. We have found that a combination of offline testing on multiple test sets (i.e. ones with implicit and sampled negatives) and K-nearest neighbor based evaluations serve as better indicators.

\subsubsection{Offline Metrics}

When training the model as a classification task we measure standard metrics on the test set such as ROC-AUC. 

In the product surfaces outlined in Section~\ref{product}, recommendations are made independently across queries. In such cases, we mainly care about the relative ordering of candidates within a given query (e.g. the order in which we show items to the user). For this reason we also measure performance using average ROC-AUC per-query. This has the added benefit of being applicable to non-classification models such as those that use triplet loss (see Section~\ref{triplet}). 
The large size of the corpus makes manually labeling evaluation data infeasible. Owing to this, we only evaluate these metrics on production data (relying on implicit and explicit user feedback). We attempted to measure recall@k by assuming unlabeled data was negative but this did not work well in practice: good models tended to retrieve good candidates from the corpus that were never shown to the user.  

Finally, in model-based candidate generation, since the majority of our training set often comes from items that were never served in production, it sometimes doesn't make sense to evaluate using per-query metrics, in which case we use per-user metrics.

\subsubsection{K-Nearest Neighbor based Evaluation Methods}  \label{knn}

While performance metrics on the test set are important we also need a stopgap evaluation to detect when performance on our test set is not reliable. These are situations where the model performs well on evaluations metrics, such as  per-user ROC-AUC, but does not retrieve good candidates.  In Section~\ref{implicit} we train such a model: It performs well on the test set but retrieves irrelevant candidates. 

K-nearest neighbor based evaluations are useful in detecting these sorts of situations. We use the trained model to generate embeddings for a sample of queries $Q'$ and candidates $C'$. Using exhaustive K-nearest neighbor exhaustive search, we generate a candidate set $C'_{q_{j}}$ from $C'$ for every query $q_j$ in $Q'$. 

Of course, we cannot measure $r_{ij}$ for most of the retrieved candidates $c_i$ in $C'_{q_{j}}$. However, we can use other coarse measures of how related $c_i$ and $q_j$ are. 

These are proxy metrics that can serve as a sanity test, and primarily act as a placeholder for qualitative evaluation, e.g. measuring how many of the retrieved candidates are a different language as the user. 

Note that a significant percentage of users consume content in a different language than their primary language or incorrectly set their primary language, so 0\% is an unrealistic and undesirable score, but a very high language mismatch percentage is indicative of poor model performance. 

Note also that if we simply wanted our model to recommend content that was always in the same language as the user, we could easily implement that with a hard constraint. However, the point of the language mismatch evaluation is that we can detect models that recommend extremely irrelevant content that would actively harm the user experience. It serves as a quantitative proxy for qualitative testing by human crowd raters.

Other times, K-nearest neighbor metrics can be used as a diagnostic criteria rather than as a metric. For example, when working on popularity corrections, we compute the Popular-Recommendation Pearson Correlation: the Pearson correlation between item popularity and its likelihood of being in users' top 5 generated recommendations. Item popularity is defined in a product-area specific way: e.g. for "who to follow" (account recommendations), it is measured as the number of followers an account has, whereas for e.g. Tweet recommendation in the home timeline it is measured as the number of engagements the Tweet has received.

\section{Techniques Part 1: Sampled Negatives} \label{solutions}

This section explains key insights we learned when training our two tower model with randomly sampled negatives. These lessons were gathered through multiple efforts across many product surfaces.  

\subsection{Sampling Ratio} \label{ratio}


In what proportion should we add randomly selected negative examples to our dataset?  We have found that with fewer sampled negative examples a model tends only to learn very broad decision boundaries. Indeed, model performance tends to increase with the number of negative samples.



\begin{figure}[hbt]
    \centering
    \includegraphics[width=0.9\columnwidth]{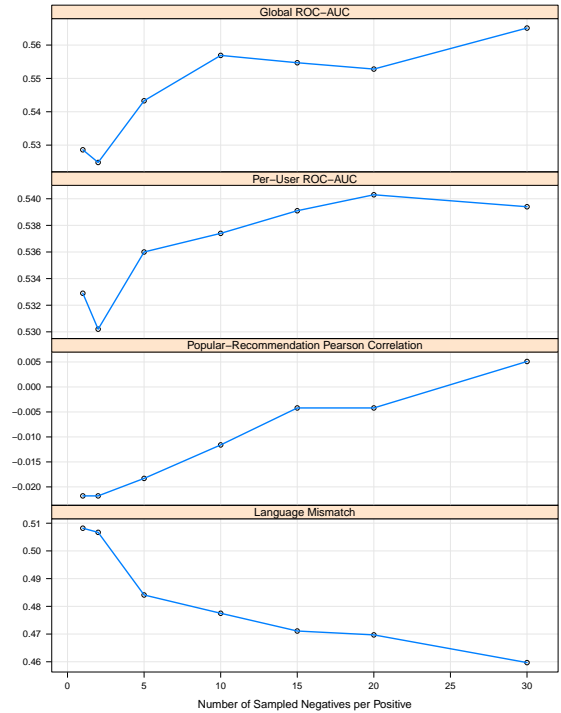}
    \caption{This plot shows how offline metrics change as we increase the number of negatives that we sample for each positive. Metric definitions described in Section \ref{evaluation}.}
    \label{fig:varying_positive_rate}
\end{figure}

To illustrate this we conduct the following experiment on a system that recommends Tweets related to other Tweets (the query and candidate are both Tweets): we train several two tower models while increasing the ratio of negatives to positives and measure offline performance. Figure~\ref{fig:varying_positive_rate} shows that offline metrics generally improve as we begin to introduce more sampled negative examples into the dataset.


\subsection{Deep Triplet Loss} \label{triplet}

Increasing the number of randomly sampled negatives lead us to explore using a comparative modeling task; specifically deep triplet loss. This is a common technique when learning embeddings and proved very successful in facial recognition tasks \cite{DBLP:journals/corr/SchroffKP15}. As far as we can ascertain, our application of it to a candidate generation recommender system is novel. Here the query $q$ is treated as an anchor and the triplet is completed with a positive and negative candidate, $c^+$ and $c^-$ respectively. The model learns to make $c^+$  closer to the anchor relative to $c$ by minimizing the loss function \begin{equation}\max (0, |q_e, c^+_e|_d - |q_e, c^-_e|_d + \alpha)\end{equation}


In negative sampling, the (query, candidate) pair has no label. As a consequence, we have to treat each one as equally irrelevant by setting $r_{ij}=0$. Note that this is a false assumption: A Tweet about baseball will not be relevant (ostensibly) to a football fanatic but is still more relevant than one about cooking. This makes it difficult for a pointwise model since we are explicitly encoding this false assumption into the task. 

The advantage of using a comparative model, such as one that uses triplet loss, is that we don’t need to specify how irrelevant a sampled candidate is. Given the triplet $(q_i, c^+, c^-)$, we only state that the negative candidate $c^-$ is less relevant than the positive candidate.  




Qualitatively, models trained with triplet loss tend to perform better. They are especially better at ranking relevant candidates above only somewhat relevant ones. 

To illustrate this we train a two-tower model using both triplet loss and cross-entropy loss. In both cases we source positive candidates, $c_i$, from instances where a user followed a suggested account. Negatives, $c_k$, were randomly sampled at a rate of 4000 negative candidates for every positive one.

We evaluate both models on a test set that uses implicit negatives. The triplet loss model sees a 17\% improvement in per user ROC-AUC over the pointwise model. 

\subsection{Popularity Correction}  \label{pop}

For each positive (query, candidate) pair $(q_j, c_i)$ in our dataset, we will sample corresponding negative samples by sampling other candidates $c_k$ according to some distribution. One option is uniformly sampling $c_k$ from $C$, where the probability of sampling any candidate $p(c_k) = 1/|C|$, and another is frequency weighting, where we sample $c_k$ in proportion to its frequency in the served traffic.

The more popular a candidate is, the more it will appear in the dataset as part of a positive pair. The popularity of candidates tends to follow a power law distribution where most content is unpopular. If we sample negatives with a uniform probability we are mostly sampling unpopular content as negatives while using mostly popular content as positives. The model can take advantage of this mismatch to achieve very high training performance by recommending the same very popular content to everyone. However, from a product standpoint, only recommending popular content (recommending it only in proportion to popularity) is undesirable, and could be achieved with a much simpler model.


Alternately, by sampling candidates in proportion to served traffic logs (not the corpus) we are in effect sampling negatives in proportion to a candidate's popularity. Sampling popular candidates as negatives leads the model to place much less importance on popularity in its recommendations, which results in less popular items being recommended. This approach is intuitively appealing, but in this extreme form the engagement rates on its recommendations tend to be much lower than on the more popular items recommended by uniform sampling.




To trade off between uniform and frequency weighting in an attempt to get the best of both worlds (we would like to recommend less globally popular content without sacrificing too much engagement), we sample negatives according to frequency and use weighting to correct for the bias in sampling popular items. Doing this allows us a degree of control between the two aforementioned unacceptable states by varying the weighting. 

To illustrate consider the following experiment.  We train a two tower model to recommend Twitter accounts to follow using deep triplet loss. For a given user and context $q_k$, positive candidates are sourced from suggested accounts she chooses to follow $a^+$. Negatives are created by sampling from every other account in the batch $a^-$. The resulting triplets $(q_k, a^+, a^-)$ were weighed in proportion to the popularity of $a^-$.  Popularity is measured here by the number of users that follow the account $\theta$ and we have found this a reasonable proxy for frequency in the dataset. 

The formula for the weights was $0.5 + 0.5 e ^ {- \theta/t}$. Note that this bounds the weights between 0.5 and 1 which we found works well in practice. Secondly, note that weight decays slower as $t$ increases.



\begin{figure}[hbt]
    \centering
    \includegraphics[width=0.9\linewidth]{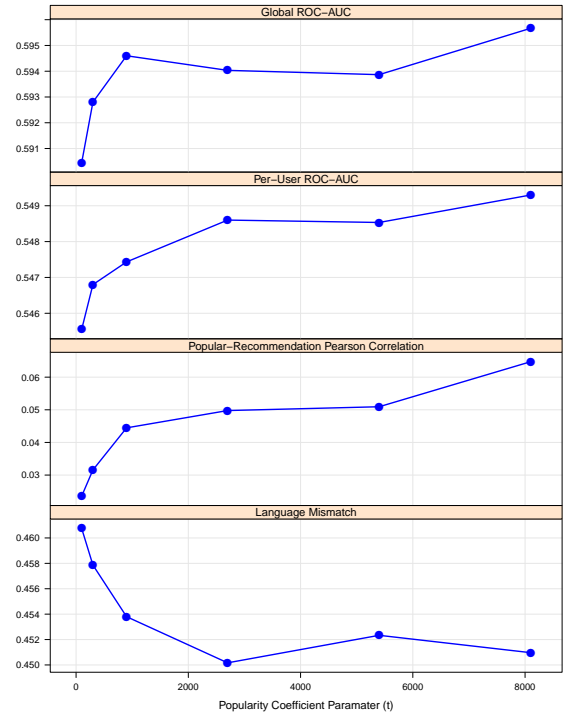}
    \caption{Model metrics improve as we increase the weighting of more popular items by increasing $t$. Metric definitions described in Section \ref{evaluation}.}
    \label{fig:correlation_t}
\end{figure}

Figure~\ref{fig:correlation_t} shows how as we increase the popularity coefficient $t$ (which increases the weight on more popular items), more popular accounts are getting recommended as evidenced by the Popular-Recommendation Pearson Correlation going up. Simultaneously, both AUC and language mismatch metrics improve. 

However, this is unsatisfying, as if we simply maximize $t$, then we would only recommend the most popular items, which we could do with a much more simple heuristic than a two-tower network. Indeed, all else equal, we have a product desire to recommend less popular items since more-popular items are very likely to be already generated by existing simple heuristic candidate generation sources. We select $t$ at the elbow where we achieve most of the gains in AUC and language mismatch while still recommending relatively unpopular accounts, which in Figure~\ref{fig:correlation_t} would be around $t=2500$.

\subsection{Online Results}

In some product surfaces a combination of all the above techniques were enough to see performance gains over the production system, such as on the "Notification Landing Page": in an online A/B test, our candidate generation system saw a relative $10.87\%$ increase in engagement using all of the above techniques when compared to the production candidate generation system, which retrieves candidates from similar authors. However, for other product surfaces, such as the home timeline, these techniques alone were not enough to see performance gains. In Section \ref{fine-tuning} we describe how we were able to beat the production system by using fine-tuning.

\section{Techniques Part 2: Fine-tuning}  \label{fine-tuning}

A problem with training a model using only randomly sampled negatives is that the task is often too easy. A trained model will learn broad decision boundaries such as language, regions, and interest categories. For this reason, the retrieved candidates tend to be only somewhat relevant to the user and most have $r_{ij}=0$ ("relevant but not engaging").

Ideally, we want to make the retrieved candidates "relevant and engaging". Doing so will make the task of the ranking phase easier, and allow the ranker to learn finer-grained details about user behavior. Served data contains exactly the data relevant to this task: almost no candidates served to the user are "extremely irrelevant", but the engaged candidates will be the most relevant. 

Inspired by work done in transfer learning, we incorporate implicit negatives into our model using fine-tuning. Transfer learning is a popular technique in several domains and has been especially successful in computer vision \cite{DBLP:journals/corr/YosinskiCBL14}. Chen et al. \cite{Chen2019AnEA} use transfer learning in the context of a socially aware recommender system by transferring knowledge between the item and social domains. 

Fine-tuning is a specific technique in transfer learning where a network pre-trained on the source task is then trained on the target task. In Howard et al. \cite{DBLP:journals/corr/abs-1801-06146} they point out that aggressive fine-tuning can cause catastrophic forgetting of the desired knowledge to be transferred. To prevent this, various techniques such as freezing hidden layers and modifying learning rates have been proposed. Our fine-tuning techniques leverages many of the existing techniques in this field. 

In this case, we first train the model on sampled random negatives, followed by fine-tuning this pre-trained model on implicit negatives. The goal of this second task is to retain the pre-trained model's ability to discriminate broad decision boundaries (such as language), while learning to discern between "relevant and engaging" and "relevant but not engaging" content.

\begin{figure}[hbt]
    \centering
    \includegraphics[width=\linewidth]{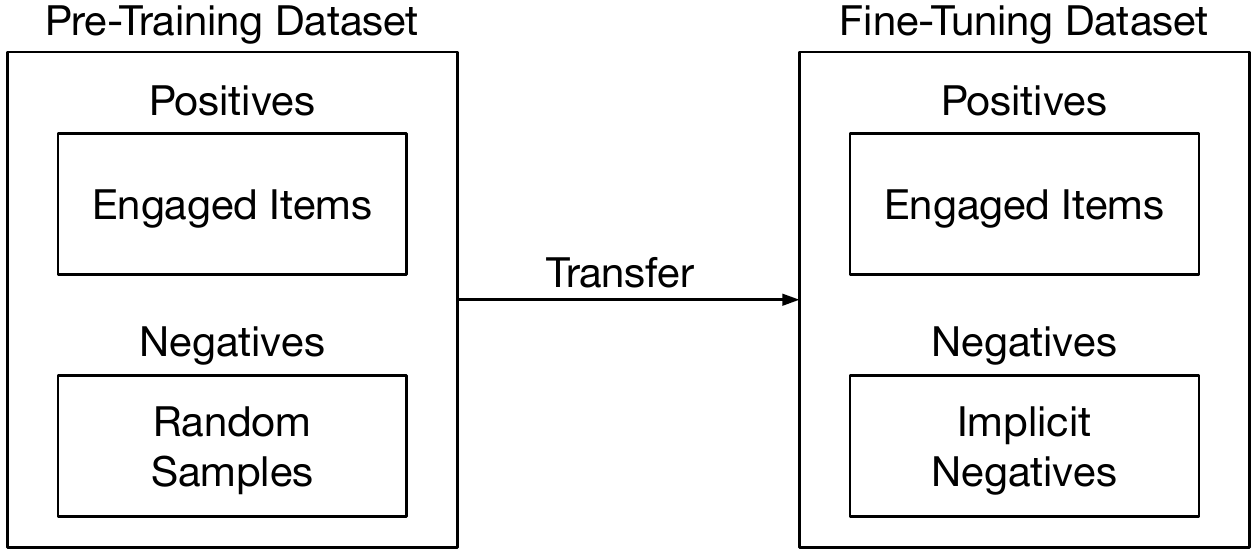}
    \caption{Fine tuning}
    \label{fig:fine_tuning}
\end{figure}

\subsection{Baseline Models}

We compare a fine-tuning model against two baseline models. One model is trained on implicit negatives from served data and the other uses randomly sampled negatives. All three models are trained for the home timeline using the setup described in Section~\ref{setup}.

\subsubsection{Implicit Feedback Model}  \label{implicit}

In this model, if a user $q_j$ engaged with candidate item $c_i$, we set $r_{ij}$ to 1. If she did not engage with $c_i$ at all we set $r_{ij}$ to 0. 

Despite the noisiness of implicit negatives, the model performs reasonably on the test set with a per-user ROC-AUC of $0.68$ and a global ROC-AUC of $0.89$. Yet when we query this model, almost all of the recommended Tweets are "extremely irrelevant", e.g. in a different language than the user: for $76\%$ of users, their top candidate (the closest Tweet neighbor) was in a different language. 

There are two important takeaways from this result. The first is that a model trained on served traffic retrieves many "extremely irrelevant" candidates. The second is that performance metrics computed over served traffic alone are misleading. As such, it's important to evaluate the model in other ways such as qualitative evaluation and using K-nearest neighbor based metrics.

Why is training on served data not enough? As mentioned in Section~\ref{problem}, the underlying issue here is one of training-serving skew. The model is being trained to discriminate on only those candidates that are somewhat and very relevant. However, during inference, the ANN compares the trained query embeddings against every candidate embedding including very irrelevant ones. 
The issue here is largely circular: We want to train a model to create a candidate set from the corpus but we only have training data that has been sourced from various candidate sets. The previous ranking systems determine our training data.

\subsubsection{Random Sampling Negative Examples} \label{random}

In this model, as before, if a user in a context $q_j$ engaged with candidate item $c_i$ we set $r_{ij}$ to 1. However, instead of using implicit negative feedback, we sample a candidate $c_k$ from the dataset and set $r_{kj}$ to 0 for the pair $q_j, c_k$.

When queried, the model does a much better job retrieving candidates that are relevant. For example, the percent of users whose top candidate was in a different language dropped to 34\%.

However, when evaluated on the test set from Section~\ref{implicit} with implicit negatives, the model only achieved a ROC-AUC of 0.54 and a per-user ROC-AUC of 0.55. This indicated the model performs nearly randomly in its ability to discriminate what content would be engaged out of the set of served candidates $C_q^s$.

These concerns are supported in an online A/B test. When compared to the current production system the percentage of engaged items (i.e. Tweets) fell by $5.5\%$.  See Section~\ref{results} for more details.

\subsection{Fine-tuned Model} \label{fine-model}

This model fine-tunes the model trained in Section~\ref{random} (with randomly sampled negatives) using the dataset outlined in Section~\ref{implicit} (with implicit negatives). We leverage several of the existing techniques outlined at the beginning of this section.

Note that we use a learning rate that is one tenth of the pre-trained model's and only trained for a single epoch. Despite this, the model had a ROC-AUC of 0.88 which is nearly identical to the one trained in \ref{implicit}. The model saw $46\%$ of top recommendations being in another language which is both low enough to serve in production and far lower than the $76\%$ mismatch of the model in Section~\ref{implicit}. This provides further evidence that per-user ROC-AUC is not reliable as a metric without also considering others such as K-nearest neighbor metrics like language mismatch.

Online tests show a significant increase in performance over the current production system: the percentage of favorited Tweets rose by a relative $3.6\%$. See Section~\ref{results} for more details. 

\subsubsection{Experimental Results} \label{results}

We test the three models presented in this section: one trained on negatives sampled from served data (Section~\ref{implicit}), another trained on randomly sampled candidates (Section~\ref{random}), and finally one that is based on fine-tuning (Section~\ref{fine-model}).

Table~\ref{tab:offline} shows the results of all three models on a dataset sampled from served data using both global and per-user ROC-AUC. We also include a K-nearest neighbors metric (Section~\ref{knn}) that tracks the percentage of language mismatches in the top retrieved candidate. 

    




\begin{table}
  \centering
  \begin{tabular}{|c|c|c|c|}
    \hline
    Model & ROC-AUC & Per-User ROC-AUC  & \thead{Language \\ Mismatch}  \\ 
    \hline
    Implicit Negatives & 0.678  & 0.885  & 0.76\(\%\)  \\ 
    \hline
    Random Negatives & 0.547 & 0.545  & 0.34\(\%\) \\ 
    \hline
    Fine-tuned & 0.670 & 0.880 & 0.46\(\%\)  \\
    \hline
  \end{tabular}
  \caption{Offline performance on a test sourced from served data.}
  \label{tab:offline}
\end{table}
\begin{table}
 \centering
 \begin{tabular}{|c|c|c|}
 \hline    
  Model & Favorite & Retweet  \\ 
 \hline
  Implicit Negatives & N/A  & N/A   \\ 
  \hline
  Random Negatives & -5.5\(\%\) & -12.5\(\%\)  \\ 
  \hline
  Fine-tuned & +3.6 \(\%\) & +2.4\(\%\)  \\
  \hline
 \end{tabular}
 \caption{Online Performance: Relative change in engagements rates. The quality of the model trained with only implicit negatives was too poor to ship online and show to users.}
 \label{tab:online}
\end{table}

Our results indicate that the fine-tuned model performs nearly as well as the implicit model on the test set (indeed it was fine-tuned on it), while maintaining nearly equal performance on language metrics. 

All of our live experiments have a similar setup. In the test group we retrieve 200 Tweets from the ANN system using the model to be tested. Those candidates are then ranked using a full ranking model. The top ranked Tweets after the ranking phase replaced Tweets in specific slots on the user's timeline designated for recommended Tweets, and the reported metrics consider only those recommended Tweet slots.

Table~\ref{tab:online} shows the results of our online A/B tests. Note that we were not able to test the implicit model in a live experiment because the quality of the retrieved candidates are deemed too poor. The fine-tuned model is able to statistically significantly improve engagements (both favorites and Retweets) over the baseline production candidate generators.

\section{Conclusion}
In this paper we presented the dataset bias problem that prevented us from training a candidate generation model on served data. We motivated the use of sampled random negatives and fine-tuning on served data.

We also presented several key insights that we have found helped improve the quality of retrieved candidates: the importance of using many random negative examples, using a comparative loss and carefully choosing the sampling distribution.

\section{Acknowledgements}
We would like everyone at Twitter that collaborated with us in the completion of this work. While naming everyone who helped us is not practical, we would like to thank Stephen Ragain, Abhishek Tayal, Yury Malkov, and Max Hansmire.

\bibliographystyle{plain}
\bibliography{references}

\end{document}